\documentstyle[12pt,epsfig]{article}
\author{Juli\'an Candia$^{a}$ and Ezequiel V. Albano$^{b}$\\
$^a${\it Departamento de F\'{\i}sica, UNLP, 
CC67,}\\{\it 1900 La Plata, Argentina}\\
$^b${\it Instituto de Investigaciones Fisicoqu\'{\i}micas
Te\'{o}ricas y Aplicadas}\\{\it (INIFTA), UNLP, CONICET, 
Suc.4, CC16,}\\{\it
1900 La Plata, Argentina}}

\title{Interfacial phase transitions in a
far-from-equilibrium magnetic growth model}
\begin{document}
\maketitle

\begin{abstract}
The irreversible growth of a magnetic film with spins having
two possible orientations
is studied in three-dimensional confined geometries of size 
$L \times L \times M$, where $M \gg L$ is the growing direction. 
A competing situation with two opposite short range   
surface magnetic fields $H$ of the same magnitude 
is analyzed. 
Due to the antisymmetric condition considered, an interface
between domains with spins having opposite orientations develops 
along the growing direction. Such interface undergoes a 
localization-delocalization transition that is the precursor 
of a wetting transition in the thermodynamic limit, in qualitative
agreement with observations performed under
equilibrium conditions. However, in contrast to its equilibrium 
counterparts, the film also exhibits a growing  
interface that undergoes a concave-convex transition 
in the growth mode. 
The phase diagram on the $H$ {\it vs} $T$ plane is
firstly obtained for a finite system, and exhibits eight different
regions. Subsequently, the phase diagram corresponding 
to the thermodynamic limit is obtained by extrapolation. It is
shown that in the latter only six regions remain.
The relevant physical properties of all these regions are
discussed in detail. 
\end{abstract}

\section{Introduction}

The interaction of a saturated gas
in contact with a wall or a substrate may result in the
occurrence of very interesting wetting phenomena,
where a macroscopically thick liquid layer condenses
at the wall, while the bulk fluid may remain in the gaseous 
phase \cite{degene,sull,diet,for,parry}. 
The wetting of solid surfaces by a
fluid is a phenomenon of primary importance in many fields
of practical technological applications (lubrication, efficiency 
of detergents, oil recovery in porous material, stability of 
paint coatings, interaction of macromolecules with interfaces, etc.
\cite{degene}). Surface enrichment or wetting layers have been
observed experimentally in a great variety of systems, such as 
e.g. polymer mixtures \cite{poly1,poly2,poly3}, 
adsorption of simple gases
on alkali metal surfaces \cite{alk1,alk2,alk3}, with the recent
addition of Hg to the adsorption species exhibiting
this kind of transition phenomena \cite{hg1,hg2}, also hydrocarbons
on mica \cite{chr}, etc. 

The study of wetting transitions at interfaces has also attracted
considerable theoretical interest \cite{diet,for,parry},
involving, among others, different approaches such as the mean 
field Ginzburg-Landau method \cite{parr,swi}, transfer matrix 
and Pfaffian techniques \cite{macio1,macio2}, density matrix 
renormalization group methods \cite{car}, solving the
Cahn-Hilliard equation \cite{ch}, using Molecular Dynamic 
simulations \cite{md}, solving self-consistent field equations
\cite{scft}, and by means of extensive Monte Carlo 
simulations \cite{eva,bind1,kur,mamu,kur1}.  

So far, the considerable progress due to all these 
studies of wetting transitions have been 
achieved for systems under {\bf equilibrium} conditions.
In contrast, the study of wetting phenomena 
under {\bf nonequilibrium} conditions has received 
much less attention.
For instance, Hinrichsen et al.\cite{hinr} have recently
introduced a nonequilibrium growth model of a one-dimensional
interface interacting with a substrate. The interface evolves
via adsorption-desorption processes, which depart from detailed
balance. Then, changing the relative rates of these processes 
a transition from a binding to a nonbinding phase 
has been reported \cite{hinr}. 

Within this context, the aim of this work is to perform
an extensive numerical study of the irreversible growth 
of a magnetic material confined between parallel walls 
where competing surface magnetic fields act. For this purpose, 
a variant of the irreversible Eden growth model \cite{eden}, 
in which particles are replaced by spins that can 
adopt two different orientations, is investigated.
It is shown that the interplay between confinement
and growth mode leads to a physically rich phase diagram.
It should also be remarked that, although the 
discussion is presented here in terms of
a magnetic language, the relevant physical concepts can 
rather straightforwardly be extended 
to other systems such as fluids, polymers, and binary mixtures.  
Apart from the fundamental interest to understand this 
complex physical situation, it may well play a key role in 
the development of technologies such as micromagnetic materials, 
micro-fluidics, self-assembly of three-dimensional structures, 
adhesives, lubricants, and coatings, among others. 
Indeed, wetting phenomena under far-from-equilibrium conditions 
are expected to be of wide
application to describe a great variety of processes 
actually encountered in practice.

Furthermore, the proposed study establishes a link with 
recent investigations of irreversible growth processes.
In fact, the study of growth systems under far-from-equilibrium
conditions is a subject that has attracted great attention 
during the last decades. Nowadays, this interdisciplinary field 
has experienced a rapid progress due to both, its interest in many 
subfields of physics, chemistry and biology, as well as by its relevance
in numerous technological applications. Recent progress in our
understanding of growth phenomena, with special emphasis on
the properties of rough interfaces, has extensively been reviewed 
\cite{fam,shl1,shl2,bar,mar}.

Also, the study of wetting phenomena in far-from-equilibrium 
systems under confinement has an extra ingredient
of theoretical interest due to the delicate interplay between
surface and bulk properties. Indeed, from the experience gained 
studying equilibrium systems,
it is well known that, using confined geometries 
with restricted dimensionality, the effects of statistical
fluctuations are more pronounced 
\cite{eva,bind1,kur,mamu,kur1,evapo,evapi,land,negro,kelv,diaz,albapa},
leading to a new and rich physical behavior, which eventually 
may be the precursor of the actual critical behavior only
observed in the thermodynamic limit.
Within this context, in the present work it is shown that
in far-from-equilibrium systems, the subtle interplay 
between finite-size effects, wetting, and interface 
growth mechanisms leads to more rich and complex
physical features than in the equilibrium counterpart.  
In fact, a complex phase diagram that exhibits a 
localization-delocalization 
transition in the interface that runs along the walls and a 
change of the curvature of the
growing interface running perpendicularly to the walls, is
evaluated and discussed, firstly for finite-size systems,
and subsequently for the extrapolated infinite system. 

This manuscript is organized as follows: 
in Section 2 we give details on the simulation method,
Section 3 is devoted to the presentation and 
discussion of the results, 
while the conclusions are finally stated in Section 4. 

\section{The model and the simulation method}

In the classical Eden model \cite{eden} on the square lattice,
the growth process starts by adding particles to the
immediate neighborhood (the perimeter) of a seed particle.
Subsequently, particles are
sticked at random to perimeter sites. This growth process
leads to the formation of compact clusters with a self-affine
interface \cite{shl1,shl2,bar,mar}.
The magnetic Eden model (MEM) \cite{mem} considers an additional
degree of freedom due to the spin of the growing particles.
Early studies of the MEM have been performed using
a single seed placed at the center of the sample \cite{mem}, but
some subsequent investigations \cite{paper1,pre,jap,prl}
have adopted instead $(d+1)-$dimensional 
rectangular geometries.
Following the latter approach, in the present work  
the MEM in $(2 + 1)-$dimensions is studied using a
rectangular geometry $L \times L \times M$ (with $M \gg L$). 
Figure 1 illustrates the general setup assumed.
The location of each site on the
lattice is specified through its rectangular coordinates $(i,j,k)$,
($1 \leq i,j \leq L$, $1 \leq k \leq M$).
The starting seed for the growing cluster is a plane
of $L \times L$ parallel-oriented spins placed at $k=1$
and cluster growth takes place
along the positive longitudinal direction (i.e., $k \geq  2$).
Periodic boundary conditions are chosen along one of the transverse 
directions (say in the $i-$direction), while 
open boundary conditions are adopted along 
the remaining transverse direction.
Competing surface magnetic fields  $H>0$  ($H'=-H$)  acting on
the sites placed at $j=1$ ($j=L$) are considered.
Then, assuming that each spin $S_{ijk}$ may adopt 
two possible orientations,
namely up and down (i.e. $S_{ijk}= \pm 1$), 
clusters are grown by selectively adding
spins to perimeter sites, which are defined as the
nearest-neighbor (NN) empty sites of the already occupied ones.
Considering a ferromagnetic interaction of 
strength $J > 0$ between NN spins, the
energy $E$ of a given configuration of spins is given by
\begin{equation}
E = - \frac{J}{2} \left( \sum_
{\langle ijk,i^{'}j^{'}k^{'} \rangle} 
S_{ijk}S_{i^{'}j^{'}k^{'}} \right) 
- H  \left( \sum_{\langle ik, \Sigma_1 \rangle } S_{i1k} -
\sum_{\langle ik, \Sigma_L \rangle } S_{iLk}  \right)  \ \ ,
\end{equation}
\noindent where $\langle ijk,i^{'}j^{'}k^{'}\rangle$
means that the summation in
the first term is taken over all occupied NN sites,
while $\langle ik, \Sigma_1 \rangle$,
$\langle ik, \Sigma_L \rangle$
denote summations carried over occupied sites on
the surfaces $\Sigma_1$,  $\Sigma_L$ (defined as
the $j=1$ and $j=L$ planes, respectively).
Thus, setting the Boltzmann constant equal to unity 
($k_{B} \equiv 1$)
and measuring absolute temperature,
energy, and magnetic fields in units of $J$, the
change of energy $\Delta E$ involved 
in the addition of a spin $S_{ijk}$
to the system is given by
\begin{equation}
\Delta E = -  S_{ijk} \left( \sum_
{\langle ijk,i^{'}j^{'}k^{'}\rangle} S_{i^{'}j^{'}k^{'}} + H
\left( \delta_{j1} -  \delta_{jL}  \right) \right) \ \ ,
\end{equation}
\noindent where the summation 
$\langle ijk,i^{ '}j^{ '}k^{ '}\rangle$
is taken over occupied NN sites keeping
$i,j,k$ fixed, 
and $\delta_{j1}$, $\delta_{jL}$ are standard Kronecker
delta symbols.
Therefore, the probability for a perimeter site
to be occupied by a spin $S_{ijk}$
is proportional to the Boltzmann factor
$\exp(- \frac{\Delta E}{T})$, where $\Delta E$
is given by Eq.(2). At each step,
the probabilities of adding up and down
spins to a given site have to be evaluated
for all perimeter sites.

\begin{figure}
\centerline{{\epsfysize=2.2in \epsffile{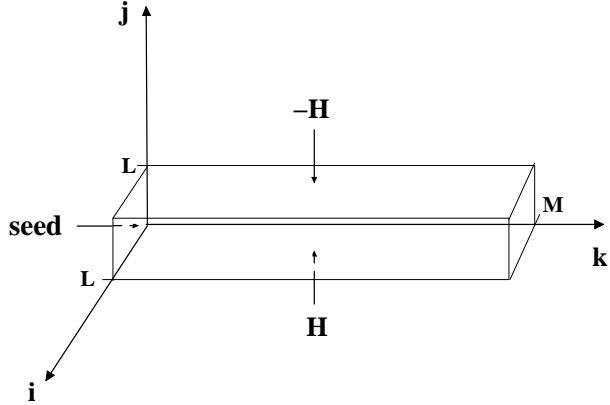}}}
\caption{General setup for the MEM in a $(2+1)-$dimensional rectangular 
geometry. The system grows along
the positive longitudinal direction from a seed constituted by
$L \times L$ parallel-oriented spins placed at $k=1$,
as indicated. Competing surface magnetic fields are applied on
the surfaces $j=1$ and $j=L$, 
while periodic boundary conditions are assumed along the 
$i-$direction. The slices shown in figure 4, 
obtained for different temperatures and magnetic fields, illustrate 
typical growth regimes.} 
\label{FIG. 1}
\end{figure}

After proper normalization of the
probabilities, the growing site and the orientation of the spin
are determined through standard Monte Carlo techniques.
Although both the interaction energy and the
Boltzmann probability distribution considered for the MEM are
similar to those used for the Ising model 
with surface magnetic fields \cite{eva},
it must be stressed that these two models
operate under extremely different conditions, namely the MEM
describes the irreversible
growth of a magnetic material and the Ising model is suitable for
the study of a magnetic system under equilibrium conditions.
In the MEM, the position and orientation of all deposited
spins remain fixed. 
The nonequilibrium nature of the MEM is clear from the fact that the
extensive thermodynamic variables (such as energy, entropy, and volume,
for instance) grow monotonically with time and tend to diverge. 
Furthermore, during the growth process, 
the system develops a rough growth interface
and evolves mainly along the longitudinal direction $k$ (see figure 1).
However, some lattice sites could remain empty even well within
the system's bulk. Since at each growth step all perimeter sites
are potential candidates for being occupied by the next spin to be added, 
these holes become gradually filled. 
It may appear that the $k$-coordinate is something like the ``time''
in a kinetic Ising model. However, this is not strictly true
because already deposited spins at position $k$ effectively
affect spin growth at position $k' < k$, and this would mean
causality violation.    

Far behind the active growth
interface, the system is compact and frozen.    
When the growing cluster interface is
close to reach the limit of the sample ($k=M$) one can
compute the relevant properties of the irreversibly frozen
cluster's bulk (in the region where the growing process has definitively
stopped), thereafter erase the useless frozen bulk, and
finally shift the growing interface 
towards the lowest possible coordinate $k$.
Hence, by repeatedly applying this procedure 
the growth process is not limited by the lattice length $M$.

It should be noticed that this paper involves a large 
computational effort. 
On the one hand, as will be seen below, the observables of interest 
(e.g. the susceptibility) are 
averaged over many transverse planes of size $L \times L$.
In order to obtain acceptably small statistical errors,
averages over $\sim 10^{5-6}$ planes in the stationary regime
are typically required. So, in the present work clusters having 
up to $\sim 10^9$ spins have been grown. On the other hand,
the update algorithm is quite slow as compared with standard
Ising simulations, since the growing probability has to be 
computed after each deposition event.

\section{Results and discussion}

Recent investigations \cite{pre} have shown that the magnetic
Eden growth process in a stripped $(d+1)-$dimensional 
geometry (with $d=1,2$) is 
characterized by an initial transient of average length $l_{Tr}$,
followed by a nonequilibrium stationary state that is 
independent of the starting
seed. It has also been shown that the MEM in $d=1$ is noncritical
(i.e., it only exhibits an ordered phase at $T=0$), 
while the MEM in $d=2$ undergoes an order-disorder thermal
transition of second order at the temperature $T_c=0.69 \pm 0.01$.
Moreover, the critical exponents associated with the continuous 
phase transition have been found to agree, within error bars, with 
those of the Ising model in two dimensions. Hence,
the reported findings have led to the conjecture that 
the $(d+1)-$dimensional MEM and the $d-$dimensional Ising 
model behave identically (unless finite-size differences 
that vanish in the thermodynamic limit) at criticality
for all $d$ \cite{pre}.

At this stage it is appropriate to briefly recall 
that a confined Ising film with competing surface fields 
undergoes an equilibrium wetting transition.
Indeed, when an Ising film is confined between two competing walls 
a distance $L$ apart from each other, 
so that the surface magnetic fields ($H$)
are of the same magnitude but opposite direction, it is found that
the competing fields cause the emergence of an interface that
undergoes a localization-delocalization transition. 
This transition shows up 
at an $L-$dependent temperature $T_{w}(L,H)$ that is 
the precursor of the true 
wetting transition temperature $T_{w}(H)$ of the infinite 
system \cite{swi,eva,bind1}.

In view of the nontrivial correspondence established between 
the $(d+1)-$dimensional MEM and the $d-$dimensional Ising model, 
it should also be expected an Ising-like wetting 
transition for the MEM. 
In fact, by applying surface magnetic fields of 
opposite signs to the MEM, 
it should be expected to obtain a well defined phase transition 
curve between wet and nonwet states on the $H-T$ plane.
In order to deal with a phase transition that will remain in the
thermodynamic limit ($L \rightarrow  \infty$), one should 
devote attention to the $(d+1)-$dimensional MEM 
with $d \geq 2$ since, 
as already pointed out, the MEM is noncritical for $d=1$.      
For this purpose, we have studied the $(2+1)-$dimensional MEM with
magnetic fields $H$ and $H'=-H$
applied to the surfaces $\Sigma_1$ and $\Sigma_L$, 
respectively (recall Section II).
As in previous investigations \cite{pre}, the mean transverse
magnetization is defined as

\begin{equation}
m \ \ (k,L,T,H) = \frac{1}{L^2} \sum_{i,j = 1}^{L} S_{ijk}                      
\end{equation}

\noindent for $k>{l_{Tr}}$, 
in order to exclude the initial transient.
Furthermore, it is assumed that the finite-size 
($L$-dependent) susceptibility
can be defined in terms of order parameter fluctuations in the 
same manner as for equilibrium systems, namely

\begin{equation}
\chi = \frac {L^{2}}{T} \left(\langle m^{2} \rangle - \langle |m| \rangle^{2} \right)
\ \ ,
\end{equation}
\noindent where $\langle ... \rangle$  
means the average taken over a
sufficiently large number of
transverse planes in the stationary regime.  
Then, using a standard procedure \cite{eva}, 
the localization-delocalization transition  
curve (on the $H-T$ plane) corresponding to the 
up-down interface running along the walls
can be obtained considering that
a point with coordinates $(H_w,T_w)$ on this curve
maximizes $\chi(H,T)$.
Figure 2 shows plots of $\chi$ {\it vs} $T$ for several values of $H$ and
the fixed lattice size $L=12$, illustrating the method used to trace
the size-dependent localization-delocalization transition curve, 
which is shown in figure 3 (open squares).
As in the case of the Ising model, this quasi-wetting 
transition refers to a transition
between a nonwet state that corresponds to a localized 
interface bound to one of the confinement walls, 
and a wet state associated to a 
delocalized domain interface centered between roughly equal domains of
up and down spins.
The localization-delocalization transition 
in a confined system is indeed the precursor of the
true wetting transition that occurs in the thermodynamic 
limit \cite{swi,eva,bind1}.
In fact, it is observed a finite jump in the wetting layer thickness
that takes place as a result of the finite size of the system.
As the lattice size is increased, the magnitude of the jump grows and
diverges in the $L\rightarrow \infty$ limit, as expected 
for a continuous wetting transition. 

\begin{figure}
\centerline{{\epsfysize=2.5in \epsffile{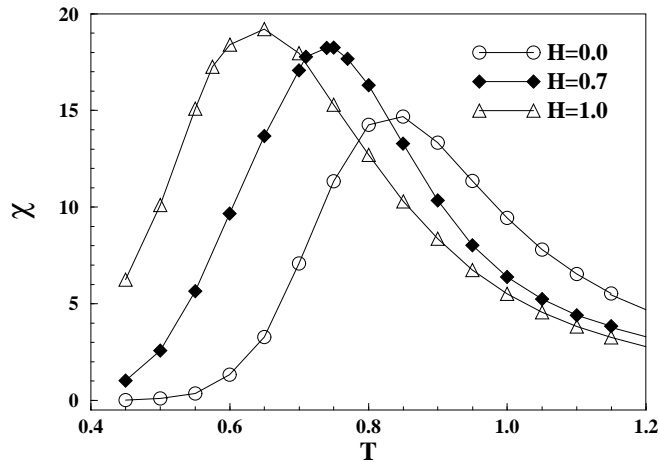}}}
\caption{Plots of $\chi$ {\it vs} $T$ for a fixed lattice 
size $L=12$ and several values of $H$,
as indicated. If $T_w$ is the temperature that 
corresponds to the maximum of $\chi$
for a given fixed value of $H=H_w$, 
then $(H_w,T_w)$ is a point on the wet-nonwet
transition curve, as follows from standard procedures [23].} 
\label{FIG. 2}
\end{figure}

Let us now discuss the critical temperature associated to the
bulk order-disorder phase transition. 
As well known from finite-size scaling theory, 
there is some degree of arbitrariness in locating  
the $L-$dependent critical temperature $T_c(L)$ of a
finite system. However, the critical point $T_c$ of the
infinite system, obtained 
by extrapolating $T_c(L)$ to the $L\rightarrow \infty$ limit,
is unique and independent of any particular choice for
the finite-size critical point.
\begin{figure}
\centerline{{\epsfxsize=4.0in \epsfysize=2.5in \epsffile{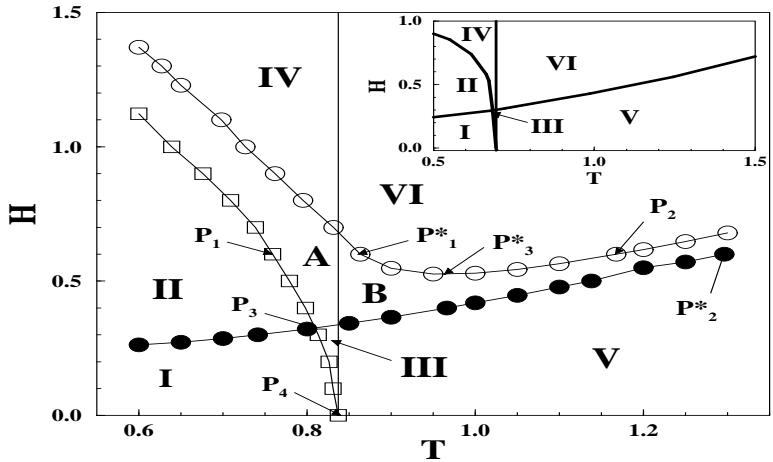}}}
\caption{$H-T$ phase diagram corresponding to a lattice of size $L=12$. 
The vertical straight line at $T_{c}(L) = 0.84$ corresponds 
to the $L-$dependent critical temperature, which
separates the low-temperature ordered phase
from the high-temperature disordered phase.
Open (filled) circles refer to the transition between non-defined and 
concave (convex) growth regimes, and squares stand for the Ising-like 
localization-delocalization transition curve.
Eight  different regions are distinguished, 
as indicated in the figure. 
Also indicated are seven representative points that are 
discussed in the text.
The inset shows the phase diagram
corresponding to the thermodynamic limit composed of
six different regions.}
\label{FIG. 3}
\end{figure}
Let us first consider the case with $H=0$, defining
the $L-$dependent critical temperature as given
by the peak of the susceptibility at zero surface field, to  
assure consistence with the evaluation 
of the localization-delocalization quasi-wetting transition.  
Indeed, under this assumption, 
the quasi-wetting curve $T_w(L,H)$ and the critical point
$T_c(L)$ coincide by definition at $H=0$.
For $L=12$, the critical point so defined is $T_c(L=12) = 0.84$,
and is shown in figure 2 by a vertical straight line.
Using larger and larger lattices, 
correspondingly smaller finite-size critical points are found, 
which tend to the actual critical point of the $(2+1)$-dimensional
MEM, namely $T_c=0.69 \pm 0.01$ \cite{pre}.
Before exploring the case $H>0$, it is appropriate to recall 
a long-standing discussion on this topic
generated in the field of equilibrium critical phenomena.
Parry and Evans \cite{parr,parr2}
claim that the critical temperature for a finite system depends
on the surface magnetic fields, and only differs from the wetting 
temperature of the infinite system ($T_w(H)$) by a term that 
vanishes in the thermodynamic limit. Indeed, they
suggest a scaling Ansatz such that $T_w(H)-T_c(L,H)$ 
is of order $L^{-1/\beta_s}$,
where $\beta_s$ is the exponent that describes the growth of the wetting layer.
However, Swift et al. \cite{swi} and Indekeu et al. \cite{indek} propose
that $T_c(L,H)$ is actually a shifted wetting transition 
(hence called quasi-wetting transition $T_w(L,H)$ ), which is 
different from the bulk critical point $T_c(L)$, 
and such that $T_w(L,H)$ tends for $L\rightarrow \infty$ to the actual wetting
temperature $T_w(H)$. In the case of the MEM, general considerations, 
supported by our numerical results, appear to favor the latter proposal, 
so that the bulk critical point is independent of $H$ and 
hence clearly different from the quasi-wetting temperature. However, 
it should be remarked that the controversy has been established for
systems under equilibrium and the present study of the MEM
corresponds to far from equilibrium conditions.

Let us first consider an increase in the surface fields from $H=0$ for a system
within the ordered phase (i.e., for $T<T_c(L)$). 
Since fluctuations in the bulk are governed only by the temperature, it 
turns clearly out that the bulk will remain in its ordered state irrespective
of the applied field. Indeed, as will be discussed below in detail, 
an increase in the fields favors the formation of a stable longitudinal
interface between domains of up and down spins. However, within each domain,
it is clear that the state of order will depend only in fluctuations
driven by the temperature. Hence, surface fields applied on an ordered system
below $T_c(L)$ will eventually favor the coexistence of oppositely oriented 
ordered domains, but are not capable of generating disorder within each domain.
These arguments are strongly supported by our simulations. For instance,
figures 4(a)-(c) show typical snapshot configurations that correspond
to nearly the same temperature (below $T_c(L=32)=0.76$) and several different 
magnetic fields. As expected from our considerations, the fields appear
to support the formation of the longitudinal interface between opposite
spin domains, but do not affect the bulk ordered state within each domain.
In particular, it should be noticed that figure 4(c) corresponds to a 
field well above the corresponding one on the quasi-wetting curve (note
that the snapshots correspond to $L=32$, and the associated transition 
curves are shifted to the left with respect to those for $L=12$, shown
in the phase diagram of figure 3). If  $T_w(L,H)$ would be
the system's critical point, it should be expected a system beyond the
quasi-wetting curve to be disordered,
in remarkable contrast with the ordered configuration shown by figure 4(c).
Moreover, as a check of consistency, one can compare the configurations
shown in figures 4(c) and 4(e), that correspond to nearly the
same fields and differ in temperature. It turns thus evident that the bulk's 
order-disorder phase transition occurs at a temperature far away
from the quasi-wetting transition curve, and consistent 
with $T_{c}(L=32) = 0.76$.

Further insights on the role of $H$ acting within the bulk 
ordered phase $T<T_c(L)$ can be gained by means of the following
procedure. Let us point our attention to the stationary regime
considering all completely filled columns directed 
along the $j-$direction, which are formed by $L$ spins 
and are identified through the values of the 
remaining coordinates $i$ and $k$. 
\begin{figure}
\centerline{{\epsfysize=3.0in \epsffile{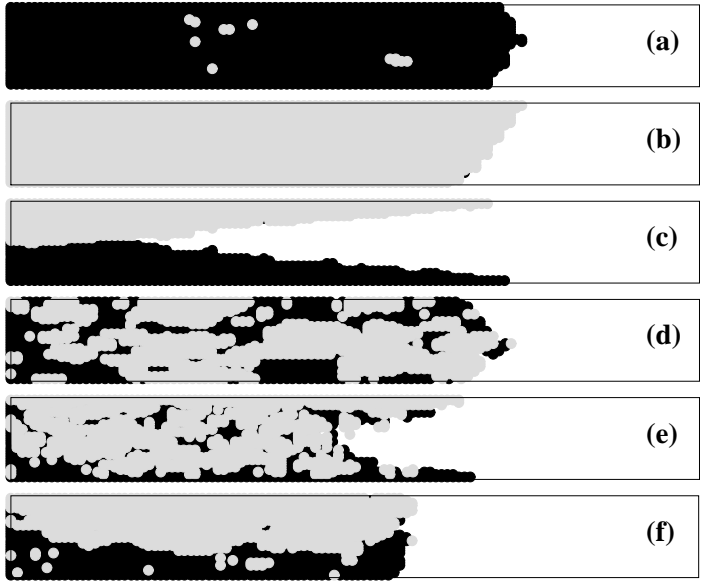}}}
\caption{Snapshot pictures showing a longitudinal slice 
given by a fixed value of the transverse coordinate $i$. 
Grey (black) points correspond to spins up (down). 
The surface field on the upper (lower) confinement
wall is positive (negative). 
The snapshots correspond to a lattice size $L=32$ and 
several different values of temperature and surface fields: 
(a)$H=0.05$, $T=0.6$; (b)$H=0.5$, $T=0.55$;
(c)$H=1.4$, $T=0.6$; (d)$H=0.1$, $T=1.0$; (e)$H=1.6$, $T=1.4$;
and (f) $H = 0.20$, $T = 0.82$.}
\label{FIG. 4}
\end{figure}
For any given column $(i,k)$, a bond to each pair of nearest-neighbor sites 
occupied by oppositely oriented spins is assigned. Summing
over the whole column, $n_b(i,k)$ is defined as the total 
number of bonds for that column, so that $n_b=0$ for 
parallel-oriented spins and $n_b=L-1$ for alternating 
up-down nearest-neighbor spins. Since all columns are 
statistically independent, the system is allow to grow 
for a sufficiently long time and averages are taken over all filled 
columns. In this way, the normalized bond probability 
distribution $P(n_b)$ can be computed as a function of temperature, 
surface magnetic fields, and lattice size. 
For the purposes of the present discussion, it suffices to fix $L=12$ and 
consider the effects of increasing the fields for a given value 
of temperature below $T_c(L=12)=0.84$.  
Figure 5 shows the bond probability distribution 
$P(n_b)$ {\it vs} $n_b$ for $T=0.6$ and
several values of $H$. It is observed that $P(n_b) \approx 0$ for $n_b \geq 2$,
irrespective of the field. Hence, it is concluded that 
the system remains in its ordered state independently of $H$, 
and that the role of the magnetic field is that of
driving the system from a state constituted by a single 
domain ($P(n_b=0) \approx 1$) to a state formed by two 
oppositely oriented ordered domains ($P(n_b=1) \approx 1$).
Figure 6 shows the mean number of bonds per column $\langle n_b \rangle$ as a 
function of $H$, for three different temperatures. As expected, in all cases 
the field drives the crossover from a single ordered domain to 
two opposite ordered domains.  

So, our discussion concerning 
the location of the critical temperature associated to 
the bulk order-disorder phase transition can be summarized 
by stating that it is found a compelling evidence for 
interpreting $T_w(L,H)$ as a quasi-wetting transition, 
clearly different from the finite-size critical
temperature $T_c(L)$. Furthermore, our results are consistent with the 
assumption of a field-independent critical point $T_c(L)$, 
since the magnetic field appears to play no role in the state 
of order within each magnetic domain. 

Since the MEM is a nonequilibrium kinetic growth model, 
it also allows the
identification of another kind of phase transition, 
namely a morphological transition associated with the curvature of the growing 
interface of the system \cite{paper1}. To avoid confusion, it shall be 
remarked that the term {\it growing interface} is used here for the 
transverse interface between occupied and empty lattice sites, 
while it was used above for the longitudinal
interface between up and down spin domains. 

\begin{figure}
\centerline{{\epsfysize=2.3in \epsffile{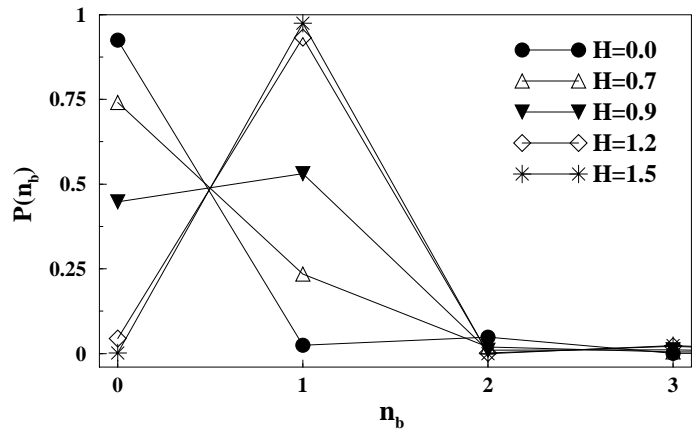}}}
\caption{Plots of the normalized bond probability distribution $P(n_b)$ vs $n_b$
for $L=12$, $T=0.6$, and several values of $H$, as indicated. 
$P(n_b)$ is negligible for $n_b > 3$, and thus it is not shown in the figure.
The role of the magnetic field appears to be that of
driving the system from a state constituted by a single domain ($P(n_b=0) \approx 1$)
to a state formed by two oppositely oriented ordered domains ($P(n_b=1) \approx 1$).}
\label{FIG. 5}
\end{figure}

\begin{figure}
\centerline{{\epsfysize=2.3in \epsffile{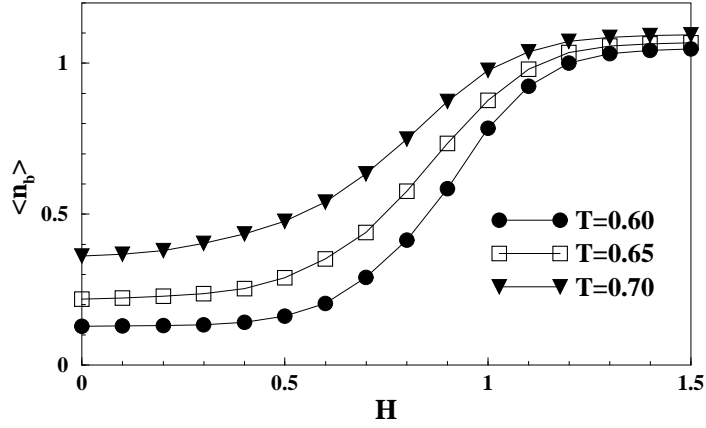}}}
\caption{Plots of $\langle n_b \rangle$ {\it vs} $H$ for a fixed lattice 
size $L=12$ and several values of $T$, as indicated.
As expected, in all cases the field drives the crossover 
from a single ordered domain to two opposite ordered domains.} 
\label{FIG. 6}
\end{figure}

Firstly, let us consider a longitudinal 
slice with a fixed value of $i$ in the range $1 \leq i \leq L$. 
In order to define the location of the growing
interface at time $t$, it is assumed that each row contributes to 
the growing interface with the outermost perimeter site 
(i.e., the site with the largest value of the 
longitudinal coordinate $k$, for a given row number $j$) 
and that number is called $I_j(t)$. 
Then, the growing interface center of mass, 
that is taken as the location of the growing interface at time $t$, $I(t)$, 
is given by

\begin{equation}
I(t) =  \frac{1}{L} \sum_{j=1}^{L} I_j(t) \ \ .
\end{equation}

Subsequently, one can evaluate the coordinates of the growing 
interface relative to its center of mass location at time $t$, 
namely $IR_j(t)  \equiv I_j(t)-I(t)$, for $j=1,2,...,L$.
In this way, it is possible to describe the growing interface at any 
time $t$ during the growing process just by evaluating the 
set $\{ IR_j(t) \}$. However, one should be cautious at this point. 
In fact, since the applied surface fields are of equal intensity 
but have opposite direction, it turns out that the probability 
of occurrence of a given growing interface $\{ IR_j \}$ must
equal the one corresponding to $\{ IR_{j^{'}} \}$, 
where $j^{'}=L+1-j$. But then, unless
the growth profile happens to be symmetric 
(i.e., invariant under $j \rightarrow  L+1-j$, for all $j$),
the time average of equally probable growing interfaces 
$\{ IR_j \}$, $\{ IR_{j^{'}} \}$ will lead to
an unphysical symmetrized profile that is not representative of 
the actual shape of the growing interface. 
To avoid this problem, the following procedure is used.
First, the largest value of the longitudinal coordinate $k$ 
that corresponds to a completely filled column is located. 
Then, by means of the sign of the total 
magnetization of that column, i.e.
$S \equiv sign \left(\sum_j S_{ijk}\right)$,
the orientation of the dominant spin domain in the active 
growing interface is identified. Supposing that, following
the recipe given above, 
a given profile $ \{ IR_j(t) \}$ is obtained,
then the growing interface location is redefined as $ \{ IR_{j^{*}}(t) \}$, 
where $j^{*} \equiv j$ if $S=+1$ and $j^{*} \equiv j^{'}=L+1-j$ 
if $S=-1$, for all $j$.
Notice that $j^{*}=1$ ($j^{*}=L$) corresponds to    
the side of dominant (non-dominant) spin domain, while
$j=1$ ($j=L$) is the side 
of positive (negative) magnetic field.

Then, it is possible to evaluate the average relative growing  
interface $\langle IR_{j^{*}} \rangle$ by taking into
account interface coordinates measured at different times 
between $t_i$ and $t_f$, and also by averaging all longitudinal 
$i-$fixed slices, i.e.

\begin{equation}
\langle IR_{j^{*}} \rangle = \frac{1}{L} \frac{1}{(t_f-t_i+1)} \sum_{i=1}^{L}
\sum_{t=t_i}^{t_f} IR_{i,j^{*}}(t) \ \ .
\end{equation} 
  
Figure 7 shows $\langle IR_{j^{*}} \rangle$ versus $j^{*}$ for different 
values of the surface magnetic field $H$, for a fixed temperature 
$T=0.6$ and a fixed lattice size $L=32$.
From the figure it follows that three qualitatively distinct growth 
regimes can clearly be distinguished. Indeed, it is observed that, 
while for small fields the system grows with convex curvature, 
increasing the fields the growth process enters into a regime of
non-defined curvature, since the dominant spin domain partially 
wets the confinement wall, while the non-dominant domain does not.
But then, further increasing the fields, a point is reached 
where the non-dominant spin domain also (partially) wets the wall and the growing 
interface turns concave. This qualitative
behavior has been observed for all temperatures and lattice 
sizes within the range of interest of this work. 

\begin{figure}
\centerline{{\epsfysize=2.5in \epsffile{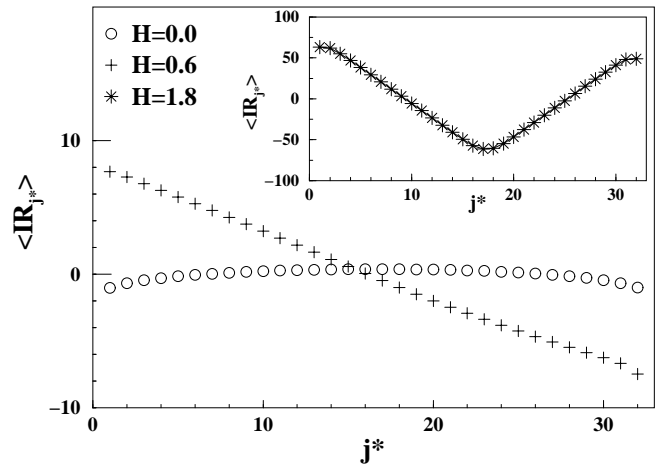}}}
\caption{Plots of the averaged interface profile 
$\langle IR_{j^{*}} \rangle$ {\it vs} $j^{*}$ for $T=0.6$ and 
different values of the surface magnetic field $H$, as indicated.
The lattice width is $L=32$. The plot corresponding to $H=1.8$
is separately shown in the inset, in order to allow a
detailed observation of the profiles for lower $H$ values.
The side $j^{*}=1$ ($j^{*}=L$) is the one corresponding to 
the dominant (non-dominant) spin domain.
Increasing the surface fields, the curvature of the growing interface changes:
convex $\rightarrow$ non-defined $\rightarrow$ concave. 
This qualitative behavior has been observed for all temperatures and 
lattice sizes within the range of interest of this work.}
\label{FIG. 7}
\end{figure}

\begin{figure}
\centerline{{\epsfysize=2.5in \epsffile{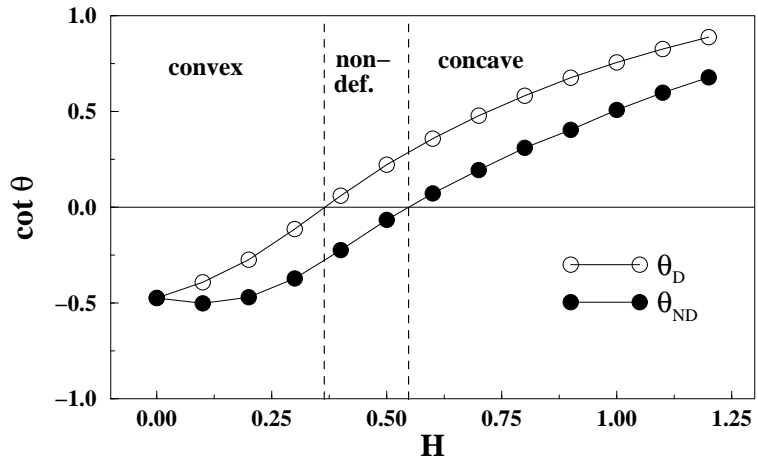}}}
\caption{Plots of $\cot(\theta)$ {\it vs} $H$ for $T=0.9$ and $L=12$.
$\theta_D$ ($\theta_{ND}$) is the contact angle corresponding to 
the dominant (non-dominant) spin cluster, and is represented 
by open (filled) circles.
The vertical dashed lines mark the fields that separate a given growth 
regime from another one, as indicated. A reference line corresponding to
$\cot(\theta)=0$ has also been included.}
\label{FIG. 8}
\end{figure}

To explore this phenomenon quantitatively, the behavior
of the contact angles between the growing interface and 
the confinement walls (as functions of temperature and magnetic field) has
to be studied thoroughly.  
Clearly, two different contact angles must be defined, namely $\theta_D$ 
for the angle corresponding to the dominant spin cluster, 
and $\theta_{ND}$ for the one that corresponds to the non-dominant
spin cluster. Figure 8 shows plots of $\cot(\theta)$ {\it vs} $H$ 
for $T=0.9$ and $L=12$. The vertical dashed
lines indicate the fields that separate a given growth regime from another one.
One observes that, increasing the surface fields, the growth regime 
changes from convex to non-defined to concave, in agreement with the 
interface profiles plotted in figure 7.   
Analogously, figures 9(a)-(d) show plots of $\cot(\theta)$ {\it vs} $T$ 
for $L=12$ and several different values for the magnetic field $H$.
Again, vertical dashed lines correspond to transition temperatures 
between different growth regimes. Figure 9(a) corresponds to $H=0.2$ 
and displays the characteristic behavior for 
very small magnetic fields, that is, a convex growing interface 
irrespective of temperature.  
For $H=0.4$ one observes a single transition from the growth 
regime of non-defined curvature to the convex growth regime, 
which shows up by increasing the temperature, 
as shown in figure 9(b). It should be noticed that the concave 
growth regime is prevented, since for small enough magnetic
fields $\cot(\theta_{ND}) < 0$ for all $T$. As the fields are increased,
$\cot(\theta_{ND})$ moves upwards and crosses 
$\cot(\theta_{ND})=0$, as expected from the plot of figure 8. 
For instance, the plots of $\cot(\theta)$ {\it vs} $T$ 
for $H=0.6$, shown in figure 9(c), exhibit this behavior. Hence, here one
has to deal with three transition temperatures. 
Finally, by further increasing the fields,
the whole low-temperature region is dominated by the concave growth 
regime and two transition temperatures remain, 
as shown in figure 9(d) for $H=1.5$.  
All these features are compactly shown on the $H-T$ phase diagram 
of figure 3, where open (filled) circles refer to the transition 
between non-defined and  concave (convex) growth regimes. 

\begin{figure}
\centerline{{\epsfysize=3.0in \epsffile{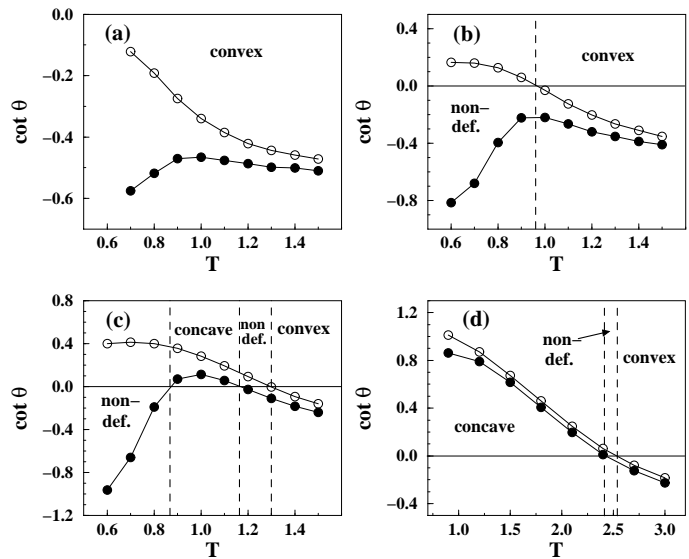}}}
\caption{Plots of $\cot(\theta)$ {\it vs} $T$ for $L=12$
and several different magnetic fields: 
(a)$H=0.2$, (b)$H=0.4$, (c)$H=0.6$, and (d)$H=1.5$.  
$\theta_D$ ($\theta_{ND}$) is the contact angle corresponding to 
the dominant (non-dominant) spin cluster, and is represented 
by open (filled) circles.
The vertical dashed lines mark the temperatures that separate a given growth 
regime from another one, as indicated. Reference lines corresponding to
$\cot(\theta)=0$ have also been included.}
\label{FIG. 9}
\end{figure}
 
As shown in figure 3, the phase diagram of the MEM in a confined geometry 
with competing surface fields is very rich and exhibits eight regions.
In order to gain some insight into the physics involved in this
phase diagram, some typical snapshot configurations characteristic 
of the various different growth regimes observed are shown in figure 4, 
as obtained using lattices of size $L=32$. 

To begin with, let us analyze Region $I$ (see figure 3), that corresponds 
to the Ising-like nonwet state and the convex growth regime. 
In this region, temperature is low and the
system grows in an ordered state, i.e. the dominant spin domain prevails and
the deposited particles tend to have their spins all pointing in 
the same direction. Small clusters with the opposite orientation 
may appear preferably on the surface where the non-dominant 
orientation field is applied. These ``drops'' might 
grow and drive a magnetization reversal, thus changing the sign 
of the dominant domain. In fact, the formation of sequences of 
well-ordered domains are characteristic of the ordered phase of 
confined (finite-size) spin systems.
For instance, this phenomenon has already been observed in finite Ising strips
\cite{eva} and magnetic Eden thin films \cite{jap}.
Due to the fact that open boundary conditions are imposed at $j=1$ and $j=L$,
perimeter sites at the confinement walls 
experience a missing neighbor effect, that is, the number
of NN sites is lower than for the case of perimeter sites
on the bulk. Since the surface magnetic fields in this region are
too weak to compensate this effect, the system grows preferentially 
along the center of the sample as compared to the walls, and the
resulting growth interface exhibits a convex shape.
A typical snapshot configuration characteristic of Region $I$ is 
shown in figure 4(a).

Let us now consider an increase in the fields, 
such as the system may enter into Region $II$ (see figure 3).
Since the temperature is kept low, the system is still in its ordered phase and
neighboring spins grow preferably parallel-oriented. The surface fields
in this region are stronger and thus capable of compensating the missing
NN sites on the surfaces. But, since the fields on both surfaces have opposite
signs, it is found that, on the one hand, the field that 
has the same orientation as the dominant spin cluster favors the 
growth of surface spins, while on the other hand,
the sites on the surface with opposite field have a lower probability to
be chosen during the Monte Carlo growth process. Hence, 
the contact angle corresponding to
the dominant spin cluster is then $\theta_D < \frac{\pi}{2} $, while the
non-dominant is $\theta_{ND} > \frac{\pi}{2} $. 
Thus, on the disfavored side the growing interface becomes pinned
and the curvature of the growing interface is not defined. 
Figure 4(b) shows a typical snapshot corresponding to Region $II$.

Keeping $H$ fixed within Region $II$ but increasing the temperature, 
thermal noise will enable the formation of drops on the disfavored side 
that eventually may nucleate into larger clusters as the temperature 
is increased even further. This process may lead to the emergence of 
an up-down interface, separating oppositely oriented domains, running in the
longitudinal direction (i.e. parallel to the walls). Since sites along
the up-down interface are surrounded by oppositely oriented NN spins, 
they have a low growing probability. 
So, in this case the system grows preferably along the confinement 
walls and the growing interface is concave (figure 4(c)). 
Then, as the temperature is
increased, the system crosses to Region $A$ (see figure 3) and 
the onset of two competitive growth regimes is observed, namely: 
{\it (i)} one exhibiting a non-defined growing curvature that appears when 
a dominant spin orientation 
is present, as in the case shown in figure 4(b); 
{\it (ii)} another that appears
when an up-down interface is established and the system
has a concave growth interface, as is shown in figure 4(c).
Further increasing the temperature and for large enough fields, 
the formation of a stable longitudinal up-down interface
that pushes back the growing interface is observed. So,
the system adopts the concave growth regime 
(see figure 4(c) corresponding to Region $IV$ in figure 3). 
Increasing the temperature beyond $T_c(L)$,
a transition from a low-temperature ordered 
state (Region $IV$) to a high-temperature disordered state
(Region $VI$, see figure 4(e)), both within the concave 
growth regime, is observed. Analogously, for small enough fields, 
a temperature increase drives the system from the ordered convex growth
regime (Region $I$) to the disordered convex growth regime (Region $V$,
see figure 4(d)). As shown in figure 3, there is also an 
intermediate fluctuating state (Region $B$) between Regions $V$ and $VI$, 
characterized by the competition between
the disordered convex growth regime and the disordered concave one.  

Finally, a quite unstable and small region (Region $III$) that
exhibits the interplay among the growth regimes of the contiguous
regions, can also be identified. Since the width of Region $III$
is of the order of the rounding observed in $T_{c}(L)$, large
fluctuations between ordered and disordered states are observed,
as well as from growth regimes of non-defined curvature to  
convex ones. However, figure 4(f) shows a snapshot configuration 
that is the fingerprint of Region $III$, that may prevail in the 
thermodynamic limit, namely a well defined spin up-down interface 
with an almost flat growing interface. 

Let us now extrapolate our results to
show that the rich variety of phenomena
found in a confined geometry is still present in the 
thermodynamic limit ($L \rightarrow  \infty$),
leading to the phase diagram shown in the inset of figure 3.
As clearly seen by comparison with the finite-size results, 
the crossover Regions $A$ and $B$ collapse in this limit,
so that only the six regions that correspond to well identified
growth regimes (as illustrated by the snapshot configurations of figure 4)
appear to remain.
  
In order to illustrate the extrapolation procedure, 
the following seven representative points of the
finite-size phase diagram are discussed in detail: 
{\it (i)} the points labeled $P_1$, $P_1^*$, $P_2$, 
and $P_2^*$, that correspond to the
intersections of the $H = 0.6$ line with the various transition
curves shown in figure 3, and {\it (ii)} the points labeled
$P_3$, $P_3^*$, and $P_4$, that refer to 
the intersection point between Regions 
$I, II, III$, and $A$, the minimum of the limiting curve between 
Regions $IV$-$VI$ and $A$-$B$, and 
the zero-field transition point, respectively.

\begin{figure}
\centerline{{\epsfxsize=4.0in\epsfysize=2.5in \epsffile{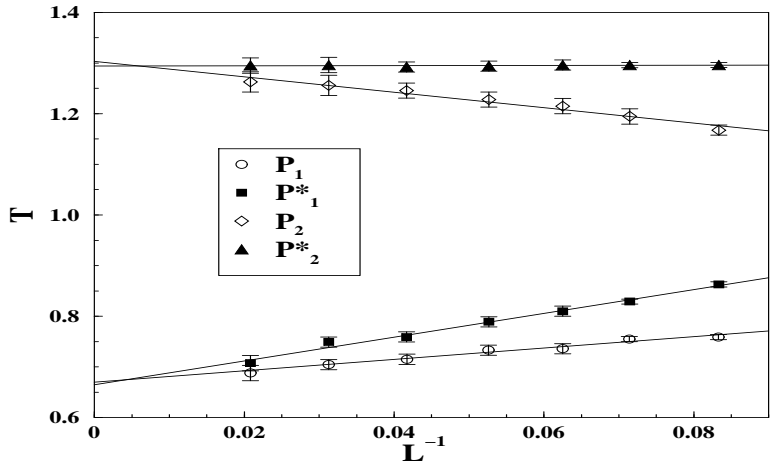}}}
\caption{Plots of $T$ versus $L^{-1}$ for $12 \leq L \leq 48$,
corresponding to the points $P_1,P_1^*,P_2$, and $P_2^*$, 
all of them with $H = 0.6$. 
The fits to the data (solid lines) show 
that, within error bars, $P_i \rightarrow P_i^*$  $(i=1,2)$ for
$L \rightarrow \infty$.}
\label{FIG. 10}
\end{figure}

Figure 10 shows plots of $T$ versus $L^{-1}$ for $12 \leq L \leq 48$ 
corresponding to the points $P_1,P_1^*,P_2,$ and $P_2^*$. 
Also shown in the figure are the fits to
the data extrapolated to $L^{-1}=0$.
The results from the extrapolations are: 
$T_1=0.67\pm0.01$, $T_1^*=0.66\pm0.01,$ and
$ T_2=1.30\pm0.02$, $T_2^*=1.29\pm0.01,$
pointing out that, within error bars, 
$P_i \rightarrow P_i^*$  $(i=1,2)$ in the $L \rightarrow \infty$
limit. Using the same procedure,
the extrapolations of $P_3$ and $P_3^*$
(not shown here) give:
$H_3=0.30\pm0.01$, $H_3^*=0.31\pm0.02,$
and $T_3=0.69\pm0.01$, $T_3^*=0.71\pm0.03$.
So, one has $P_3 \rightarrow P_3^*$ for $L \rightarrow \infty$  
within error bars. Finally, the extrapolation of $P_4$ is
$T_4 =  T_c = 0.69\pm0.01$.
  
Using the above-mentioned extrapolation procedure,
the phase diagram in the thermodynamic limit can be drawn,  
as shown in the inset of figure 3. 
By comparison with the finite-size phase diagram of
figure 3, one can notice that, as anticipated, 
the crossover Regions $A$ and $B$ appear in the
phase diagram just as a consequence of the finite-size 
nature of confined geometries, since they collapse in the
$L \rightarrow \infty$ limit.
Moreover, we conjecture that Region $III$ may remain
in the thermodynamic limit. 
Although this (very tiny!) region corresponds to
a physically well characterized growth regime, since
one expects that the system in this region may grow in an ordered phase
with a delocalized up-down domain interface and a convex growing interface,
statistical errors due to large fluctuations close to criticality 
hinder a more accurate location of this region. The unambiguous
clarification of our conjecture remains as an open question
that will require a huge computational effort. 

Besides an Ising-like continuous wetting transition, coupled
morphological transitions in the growing interface, which arise
from the MEM's kinetic growth process, have also been identified.
Comparing the equilibrium wetting phase diagram of the Ising model
\cite{parr,eva,bind1} and that
of the MEM, it follows that the nonequilibrium
nature of the latter introduces new and rich  physical features of interest:
the nonwet (wet) Ising phase splits out into Regions $I$ and $II$
(Regions $III$ and $IV$), both within the ordered regime ($T < T_{c}$) 
but showing an additional transition in the interface growth mode. 
Also, the disordered state of the Ising system ($T > T_{c}$) splits
out into Regions $V$ and $VI$ exhibiting a transition in the 
interface growth mode. 

It should be noticed that we have restricted ourselves 
to temperatures above $T=0.5$ throughout, since the lower the 
temperature in the ordered phase, the greater the computational 
effort needed to reliably sample the whole
configuration space (indeed, ergodicity is broken in the $T \rightarrow 0$ limit).
Right at $T = 0.5$ the wetting curve of the phase diagram
(inset of figure 3) intercepts the $H$-axis close to $H = 0.9$.
On physical grounds no particular features of interest are
expected to arise in the $T \rightarrow 0$ limit, and 
the critical field $H = 1$ for $T = 0$ can be inferred
by energetic considerations, as e.g. in the case of the Ising model. 

\section{Conclusions}

The growth of magnetic Eden clusters with
ferromagnetic interactions between nearest-neighbor spins has
been studied in a $(2+1)-$dimensional geometry with competing 
surface magnetic fields.
Extensive Monte Carlo simulations allow us to locate, on the one hand, 
an Ising-like localization-delocalization wetting transition, and, 
on the other hand, a morphological transition associated with
the curvature of the growing interface. In this way, 
eight different regions on the $H-T$ phase
diagram for a finite-size lattice are identified. 
Moreover, the characteristic behavior of typical growth 
processes within each region are discussed, and 
qualitative explanations that account for the observed features
are provided. Finally, extrapolating the results to 
the $L \rightarrow  \infty$ limit the phase diagram is
obtained. It is composed of six different regions,
since two crossover regions identified in the finite-size phase diagram
appear to collapse in the thermodynamic limit. 
The obtained phase diagram
shows new and rich physical features of interest, 
which arise as a consequence of the nonequilibrium nature of 
the investigated model. 

We hope that the presented results will, on the one hand, contribute to 
the understanding of the rich and complex physical phenomena exhibited by the
irreversible growth of binary mixtures in confined
geometries, and on the other hand, stimulate 
further experimental and theoretical work.

\vskip 1.0 true cm
{\bf  ACKNOWLEDGMENTS}. This work is  supported  financially by
CONICET, UNLP, and ANPCyT (Argentina). The authors thank the referees
for fruitful comments and suggestions.

\end{document}